\documentclass[aps,amssymb,amsmath,
twocolumn,showpacs,superscriptaddress,
groupedaddress,ifmbe]{revtex4-1}

\usepackage{graphicx}
\usepackage{natbib}
\usepackage{dcolumn}
\usepackage{bm}
\usepackage[table,xcdraw]{xcolor}
\usepackage{nth}
\usepackage[ruled,vlined]{algorithm2e}
\usepackage{framed}
\usepackage{listings}
\usepackage[titletoc,title]{appendix}
\usepackage{tikz}
\usepackage{lipsum}
\usepackage{hyperref}
\usepackage{multirow}
\usepackage{nicefrac}
\usepackage[table]{xcolor}
\usepackage{threeparttable}
\usepackage{comment}
\usepackage{amsmath}
\usepackage[autostyle]{csquotes}
\usepackage{float}
\usepackage{placeins}

\hypersetup{
  colorlinks,
  citecolor=red,
  linkcolor=blue,
  urlcolor=blue}

\begin{document}
\title{Complete and Orthonormal Sets of Exponential$-$type Orbitals\\with non$-$integer quantum numbers.\\
On the results for many$-$electron atoms using Roothaan's LCAO method.}
\author{Ali Ba{\u g}c{\i}}
\email{abagci@pau.edu.tr}
\affiliation{Computational and Gravitational Physics Laboratory, Department of Physics, Faculty of Science, Pamukkale University, Denizli, Turkey}
\author{Philip E. Hoggan}
\affiliation{Institut Pascal, UMR 6602 CNRS, BP 80026, 63178 Aubiere Cedex, France.}

\begin{abstract}
Complete orthonormal sets of exponential$-$type orbitals with non$-$integer principal quantum numbers are discussed as basis functions in non$-$relativistic Hartree$–$Fock$–$Roothaan electronic structure calculations of atoms. A method is proposed to construct accurate and computationally efficient basis sets using these orbitals. It is demonstrated that principal quantum numbers of fractional order cannot be treated solely as variational parameters, since such a procedure may lead to unphysical basis sets (in particular, linearly dependent Slater$-$type functions). Ground$-$state total energies for the Be$-$ and Ne$-$ isoelectronic series are calculated. The results obtained are lower than those reported using other published basis sets. However, the energies obtained using Slater$-$type functions with non$-$integer principal quantum numbers are omitted from the comparison. These \enquote{orbitals} have no physical interpretation (except the \enquote{$1s$}, which coincides with a hydrogenlike eigenfunction). In general linear independence of such Slater$-$type orbitals is not guaranteed. The results confirm that the parameter $\alpha$, used to represent the complete orthonormal exponential$-$type orbitals in the weighted Hilbert space, is neither observable nor suitable to be considered as a variational parameter, despite its treatment as such in some prior work.
\begin{description}
\item[Keywords]
Exponential$-$type orbitals, Non$-$integer quantum numbers, Self-consistent 

field method, Multielectron atoms
\end{description}
\end{abstract}
\maketitle
\section{Introduction} \label{intro}
In the computation of electronic structures for atoms and molecules, basis functions are employed that must be square$-$integrable and capable of spanning the space of eigenfunctions associated with the corresponding Hamiltonian \cite{1_Higgins_1977}. \\
For atoms, Sturm$-$Liouville theory can be used to provide a complete orthonormal orbital basis \cite{2_Shull_1959, 3_Rotenberg_1962, 4_Rotenberg_1970}. These functions possess exponential decay, with a screening parameter that does not involve quantum numbers (contrary to the hydrogenlike functions \cite{5_Schiff_1968}, where it is $Z/n$ for nuclear charge $Z$ and principle quantum number $n$). The screening parameters are fixed by orthogonalising Coulomb repulsion. They are complete, orthonormal and include the full spectrum of continuum states \cite{6_Hylleraas_1928}.

The innovation in conception of the complete and orthonormal sets of exponential$-$type orbitals with non$-$integer quantum numbers, namely Ba{\u g}c{\i}$-$Hoggan complete and orthonormal sets of exponential$-$type orbitals (Ba{\u g}c{\i}$-$Hoggan ETOs, BH$-$ETOs) \cite{7_Bagci_2023}, described in the first article of this series \cite{8_Bagci_2025}, is that they are solutions to the Dirac equation, after passage to the non$-$relativistic limit. All Dirac's theory \cite{9_Dirac_1978} is incorporated, which also proves they are a complete orthonormal basis, or kets spanning Hilbert space
(\textit{vide infra}). Furthermore, they can be obtained by solving the Kepler problem defined by Infeld and Hull \cite{10_Infeld_1951}. The space spanned by these eigenfuncions is beyond standard Hilbert space which is a subspace of the present basis space.
In fact, it is necessary to define Hilbert$-$Sobolev spaces \cite{11_Sobolev_1963, 12_Homeier_1992}. A class of discretizations must be defined for each class of BH$-$ETOs, and a specific type of Sobolev space is spanned by these orbitals. Each space based on the $\nu$ non$-$integer parameter has its own Sobolev space. \\
As a corollary, extension of Coulomb$-$Sturmians or Shull Lambda functions \cite{13_Shull_1955} to non$-$integer quantum numbers is not defined in the space that integer $n$ functions span.

It is worth mentioning the special case of the ground state of Helium. This is of interest because the $\left(1s^2\right)$ electron configuration can be expressed equivalently in a Slater$-$type function (STF) basis or by using orbital functions such as hydrogen-like functions. In fact, in 1964 Zung and Parr \cite{14_Zung_1964} showed that the linear combination of $1s$ and $0s$ STFs approaches the Hartree$-$Fock limit (referred to below) \cite{15_Fischer_1977} for the He$-$atom ground state to within 10$\mu$ Hartree. This triggered interest in the $0s$ function, best defined in momentum space \cite{16_Robinson_1966, 17_Szalewicz_1981, 18_Simas_1983, 19_Robertson_1986}. Later, Koga \cite{20_Koga_1998} extended this wave$-$function to a hyperbolic cosine, which, however, should be reserved for helium, but was used numerically for atoms with more than two electrons.

The properties of BH$-$ETOs ensure that the variational limit approaches the Hartree$-$Fock (HF) limit from above, thereby providing a reliable approximation. Within the framework of the self-consistent field method \cite{21_Hartree_1928, 22_Fock_1930}, complete and orthonormal sets of exponential$-$type orbitals (ETOs) introduce additional variational flexibility by using non$-$integer quantum numbers \cite{7_Bagci_2023}. This approach necessitates the definition of non$-$integer order derivatives and the corresponding orthogonal polynomials. The authors have previously addressed this challenge \cite{23_Bagci_2020, 24_Bagci_2022, 25_Bagci_2024, 26_Bagci_2025} (see also references therein), showing that the resulting variational procedure remains numerically stable.
The first paper in this series \cite{8_Bagci_2025} focused on two$-$electron atomic systems. In the present work, we extend the analysis to many$-$electron atoms by employing solutions of the Hartree$-$Fock equation to evaluate the effectiveness of the ETOs formulation. Numerical computations are performed for Be$-$like and Ne$-$like atomic systems. The present study is planned as follows:
\begin{itemize}
\item To investigate whether the Hartree$-$Fock limit can be approached using ETOs.
\item To examine the convergence behavior of ETOs in atomic electronic structure calculations.
\item To present numerical results based on the orbital parameters refined by Koga \cite{27_Koga_1995, 28_Koga_1995}, which represent a systematic improvement over the earlier tabulations by Clementi, Roetti \cite{29_Clementi_1974} and Bunge et al. \cite{30_Bunge_1993}.
\end{itemize}
In all the systems studied, electrons are considered to be spin$-$paired, one with spin up and the other with spin down. The basis set is progressively expanded to include up to approximately ten functions. Variational parameters are optimized numerically by minimizing the total energy, subject to accurate evaluation of the Hamiltonian matrix elements in the chosen basis. This requires the computation of both single$-$electron and two$-$electron integrals, including closed$-$form expressions for Coulomb attraction, electron$-$electron repulsion, and exchange integrals.\\
Before presenting these results, the following section details the method for constructing an accurate basis set in which the quantum numbers are treated as parameters, thereby removing conventional restrictions and enhancing variational flexibility.
\section{On the use of Exponential$-$type basis sets in Hartree$-$Fock$-$Roothaan Equations} \label{hfrbasis}
The Hartree$–$Fock method provides matrix solution of the Schr{\"o}dinger equation, for many$-$electron atoms and molecules. The non$-$relativistic kinetic energy operator is conserved in its analytical expression and a screening$-$field approximation is introduced, in which each electron moves in an averaged potential generated by the nucleus and all other electrons. This iteratively produces the so-called Self-consistent Field (SCF) \cite{21_Hartree_1928, 22_Fock_1930}. The SCF equations resulting from the Hartree$-$Fock method are solved by means of the linear combinations of atomic orbitals (LCAO) \cite{31_Roothaan_1951}. This is an orbital approximation to the many$-$electron wave function represented by the Slater determinant \cite{32_Slater_1929}, to ensure exchange antisymmetry. Despite the presence of electron$-$electron interactions, the wave function obtained from the solution maintains radial and angular forms closely resembling hydrogenlike atomic orbitals, meaning that the radial factors in the fully orthonormal set of functions (used here and expressed via associated Laguerre polynomials) possess the characteristic radial nodes of hydrogenlike atom electron densities.\\
The selection of basis functions in the LCAO approximation is flexible; however, these functions should satisfy the cusp condition \cite{33_Kato_1957, 34_Agmon_1982} and \enquote{saturate the space} with few orbitals, giving rapid convergence of the calculated energy and wave function towards the complete basis set limit (the numerically exact Hartree$-$Fock solution, HF$-$limit). Consider the Slater$-$type basis (Slater-type Functions, STFs) \cite{35_Slater_1930},
\begin{align}\label{eq:1}
\chi_{nlm}\left(\zeta, \vec{r}\right)
=R_{n}^{n-1}\left(\zeta, r\right)S_{lm}\left(\theta, \varphi\right),
\end{align}
where,
\begin{align}\label{eq:2}
R_{n}^{n-1}\left(\zeta, r\right)
=\dfrac{\left(2\zeta \right)^{n+1/2}}{\sqrt{\left(2n\right)!}}r^{n-1}e^{-\zeta r}.
\end{align}
The functions $S_{lm}$ are normalized complex $\left(S_{lm} \equiv Y_{lm}, Y^{*}_{lm}=Y_{l-m} \right)$ \cite{36_Condon_1935} or real spherical harmonics. STFs are a simplified choice of basis in solving the  HF equations using the LCAO approximation because they are obtained by taking the highest power of $r$ in hydrogenlike orbitals. STFs are radially nodeless and non-othogonal over $r$, however. An extended class of STFs, wherein the principal quantum number is allowed to take non$-$integer values \cite{37_Zener_1930} (a hypothesis proposed by Parr and Joy \cite{38_Parr_1957}), has been considered for atomic and molecular calculations. In that case, the factorial in Eq. (\ref{eq:1}) is replaced by a gamma function.  Principal quantum numbers are not restricted to integers, aiming to enhance the flexibility of the basis set and accelerate convergence, even though the cusp condition is no longer satisfied. In pursuit of the same goal, STFs were further extended using hyperbolic functions \cite{20_Koga_1998, 39_Koga_1998, 40_Koga_1999},
\begin{multline}\label{eq:3}
R_{n}^{n-1}\left(\zeta^{}_{1},\zeta^{}_{2}, \zeta^{}_{3}, \vec{r}\right)
\\
=N_{n}\left(\zeta^{}_{1}, \zeta^{}_{2}, \zeta^{}_{3} \right)r^{n-1}e^{-\zeta^{}_{1} r}
cosh\left(\zeta^{}_{2} r+\zeta^{}_{3}\right)
\end{multline}
$N_{n}\left(\zeta^{}_{1}, \zeta^{}_{2}, \zeta^{}_{3} \right)$ are the normalization constants. Such functions (related to the STFs) were applied to atoms with multiple alternatives, the last even adding a parameter to exponential and hyperbolic functions as $e^{\zeta_{1} r^{\mu}}$, $cosh\left( \zeta_{2}r^{\zeta_{3}} + \zeta_{4} \right)$ \cite{41_Koga_1997, 42_Guseinov_2012, 43_Erturk_2012, 44_Erturk_2012}, respectively. The parameter $\mu$ was also treated variationally. Suggesting a pre$-$complex form of Slater$-$type functions with non$-$integer quantum numbers (NSTFs), is another hypothesis motivated primarily by concerns for improved convergence \cite{45_Koga_1997, 46_Koga_1997, 47_Koga_1997, 48_Koga_2000, 49_Guseinov_2008, 50_Erturk_2009}. It leads to a linear dependence and an extension of the electronic wave function within the LCAO approximation. Given that NSTFs are presumed not to satisfy any differential equation, they were regarded as independent functions (which is not the case). The expansion is thus allowed, in consequence, to have following form:
\begin{multline}\label{eq:4}
\psi_{n_{j}l_{j}m_{j}}\left(\vec{r}\right)
\\
= c_{1j}\chi_{n_{1}^{\ast} l_{1}^{} m_{1}^{}}\left(\zeta^{}_{1},\vec{r}\right)+
c_{2j}\chi_{n_{2}^{\ast} l_{2}^{} m_{2}^{} }\left(\zeta^{}_{2}, \vec{r}\right)+ ...
\end{multline}
more generally,
\begin{align}\label{eq:5}
\psi_{n_{j}l_{j}m_{j}}\left(\vec{r}\right)=
\sum_{i}c_{ij}\chi_{n_{i}^{\ast} l_{i}^{} m_{i}^{} }\left(\zeta^{}_{i}, \vec{r}\right),
\end{align}
here, $\left(\zeta_{i}, n_{i}^{}\right)$ are the variational parameters and principal quantum numbers to be variationally optimized, respectively.\\
NSTFs on the other hand, find their origin in Ba{\u g}c{\i}$-$Hoggan ETOs \cite{7_Bagci_2023}. Their radial parts are expressed as follows:
\begin{multline}\label{eq:6}
R_{n^{\ast}l^{\ast}}^{\alpha\nu}\left( \zeta, r \right)
\\
=\mathcal{N}_{n^{\ast}l^{\ast}}^{\alpha\nu}\left(\zeta\right)
\left(2\zeta r\right)^{l^{\ast}+\nu-1}e^{-\zeta r}
L_{n^{\ast}-l^{\ast}-\nu}^{2l^{\ast}+2\nu-\alpha}\left(2\zeta r\right),
\end{multline}
$\left\lbrace n^{\ast},l^{\ast} \right\rbrace \in \mathbb{R}$ and $0 < \nu \leq 1$, $\alpha$ is a weighting parameter in Hilbert space.
By taking the highest power or $r$ in the Eq. (\ref{eq:6}), the Eq. (\ref{eq:1}) with non$-$integer principal quantum numbers are obtained. Each $n_{i}^{\ast}$ in the Eq. (\ref{eq:5}) arise from the BH$-$ETOs functions,
\begin{align}\label{eq:7}
 \Psi_{n^{\ast}l^{\ast}m}: \mathbb{R}^{3}\rightarrow \mathbb{C},
\end{align}
where, $n^{\ast}=\nu+k$, $k \in \mathbb{Z}_{\geq 0}$, $\nu \in \mathbb{R}_{> 0}$.\\
The Hilbert space spanned by BH$-$ETOs is defined as,
\begin{align}\label{eq:8}
\mathcal{H}:= \bigcup_{\nu \in \mathbb{R}_{>0}} \mathcal{H}_{\nu},
\end{align}
\begin{align}\label{eq:9}
\mathcal{H}_{\nu}
=\biggl\{
f \in \mathcal{L}^{2}\left( \mathbb{R}^{3} \right)
\Bigm\lvert
f=\lim_{N\rightarrow \infty}\sum_{i=1}^{N}c_{i}\Psi_{n^{\ast}_{i} l^{}_{i} m^{}_{i}} 
\biggl\},
\end{align}
They are convergent in $\lVert .  \rVert_{\mathcal{L}^{2}}$.  Notice that the definition given in Eq. (\ref{eq:9}) is based on the structure of a basic Hilbert space. This formalism can be generalized to include inner products, norms, and function spaces involving weight functions.\\
Each term in the expansion given in Eq. (\ref{eq:5}) belongs to a distinct space $\mathcal{H}_{\nu_i}$, where $i = 1, 2, 3, ...$\hspace{1mm}. The set of functions within each $\mathcal{H}_{\nu}$ is complete and orthonormal, whereas a set of functions drawn from different $\mathcal{H}_{\nu}$ spaces is neither orthonormal nor complete. Although the expansion in Eq. (\ref{eq:5}) provides better convergence, the considerable computational effort required for intensive parameter optimization is a serious drawback regarding the viability of NSTFs in non$-$relativistic electronic structure calculations. It is generally not even possible to interpret the result physically, since this basis does not constitute a set of atomic orbitals (they are not eigenfunctions of an atomic Hamiltonian). Considering the set of non$-$integer principal quantum numbers as given in Eq. (\ref{eq:5}) together with additional complexity for integral evaluation, standard Slater$-$type functions are capable of achieving the HF$-$limit for total energy with minimal parameter optimization. BH$-$ETOs, on the other hand, introduce only one additional parameter $\left( \nu \right)$ to be optimized beyond those present in standard STFs and any basis set constructed from them. The completeness and orthonormality of BH$-$ETOs ensure that any square$-$integrable function can be expanded accurately about an arbitrarily displaced center. This property is crucial for applications involving molecular systems, where orbitals centered on different nuclei must be represented using a common basis. However, a detailed treatment of such multicenter expansions lies beyond the scope of the present paper. Note that BH$-$ETOs arise as solutions to the generalized Infeld$-$Hull Schrödinger$-$like differential equation. They provide a more accurate analytical representation of the relativistic limit of the Dirac equation \cite{9_Dirac_1978, 26_Bagci_2025} compared to Coulomb$-$Sturmian functions \cite{2_Shull_1959, 3_Rotenberg_1962, 4_Rotenberg_1970}.
\section*{Method of Computation}
A computational implementation is developed in the Mathematica programming language to solve the Hartree$–$Fock$–$Roothaan equations within the self-consistent field approximation. The Cholesky decomposition method \cite{51_Golub_2013} is used to solve Roothaan's generalized eigenvalue equation.\\
In solving the Roothaan equations of the form \cite{31_Roothaan_1951}
\begin{align} \label{eq:10}
\mathbf{F} \mathbf{C} = \mathbf{S} \mathbf{C} \boldsymbol{\varepsilon},
\end{align}
where $\mathbf{F}$ is the Fock matrix, $\mathbf{S}$ is the overlap matrix, $\mathbf{C}$ is the molecular orbital coefficient matrix, and $\boldsymbol{\varepsilon}$ is the diagonal matrix of orbital energies, one encounters a generalized eigenvalue problem due to the presence of the non$-$orthogonal basis represented by $\mathbf{S}$. Since $\mathbf{S}$ is Hermitian and positive$-$definite, it admits a Cholesky decomposition:
\begin{align} \label{eq:11}
\mathbf{S} = \mathbf{L} \mathbf{L}^\dagger,
\end{align}
where $\mathbf{L}$ is a lower triangular matrix. By defining a transformed coefficient matrix $\mathbf{C}' = \mathbf{L}^\dagger \mathbf{C}$, Roothaan's equations can be rewritten as a standard eigenvalue problem:
\begin{align} \label{eq:12}
\tilde{\mathbf{F}} \mathbf{C}' = \mathbf{C}' \boldsymbol{\varepsilon},
\end{align}
where $\tilde{\mathbf{F}} = \mathbf{L}^{-1} \mathbf{F} \mathbf{L}^{-\dagger}$.\\
This transformation eliminates the overlap matrix from the right$-$hand side, effectively orthonormalizing the basis and enabling the use of standard Hermitian eigenvalue solvers. The Cholesky approach is both numerically stable and computationally efficient, particularly advantageous in large-scale electronic structure calculations where the direct inversion of $\mathbf{S}$ may be ill$-$conditioned or costly.

The optimization of nonlinear parameters associated with the orbital parameters is performed using Powell’s method \cite{52_Powell_1964}, a derivative$-$free algorithm for unconstrained minimization in multidimensional parameter spaces. Powell’s method iteratively minimizes the objective function by performing a sequence of one$-$dimensional line searches along a set of conjugate directions. Formally, given a function $f: \mathbb{R}^n \to \mathbb{R}$, the algorithm seeks:
\begin{align} \label{eq:13}
\min_{\mathbf{x} \in \mathbb{R}^n} f(\mathbf{x}),
\end{align}
by successively solving
\begin{align} \label{eq:14}
\min_{\alpha \in \mathbb{R}} f(\mathbf{x} + \alpha \mathbf{d}_i)
\end{align}
for each direction $\mathbf{d}_i$, $i=1, \ldots, n$, where $\mathbf{d}_i$ are updated iteratively to improve convergence without requiring derivative information.\\
For each one$-$dimensional line minimization, the golden-section search method is used. This technique exploits the golden ratio of $\varphi = \frac{1+\sqrt{5}}{2} \approx 1.618$, approximately. This is used to efficiently bracket and reduce search intervals, minimizing the number of function evaluations needed to locate the minimum along a line. 
\section*{Results and Discussion}
The basis set approximations used in calculations are determined for each angular momentum quantum number $l$, according to $\left\lbrace n^{\ast} \right\rbrace_{l}= \left\lbrace p+\nu \right\rbrace_{l}$, where $p \in \mathbb{N}_{0}$ and $\nu \in \left(0,1\right]$. This symbolization aligns with conventional orbital designations: for instance, $p=0$, $l=0$ corresponds to the $1s$ orbital; $p=1$, $l=0$ to $2s$; $p=1$, $l=1$ to $2p$, etc. The notation $\left( ... - .. - . \right)$, which appears in the tables is used to represent the constructed basis set. It stands for the principal quantum number of constituent orbitals and the symbol $-$ separates different $l$ values in the order. The optimization is carried out for variational parameters $\zeta$, $\nu$, $\alpha$ for any subset of $\left\lbrace n^{\ast} \right\rbrace_{l}$. 

Numerical calculations are performed for the Be$-$ and Ne$-$isoelectronic series and for F$^{-}$ and Li$^{-}$ anions. The results for total energy are presented in Tables  \ref{tab:table1} and \ref{tab:table2}. Lists of optimized variational parameters used to obtain the results given in the Tables \ref{tab:table1} and \ref{tab:table2} are given in Tables \ref{tab:table3} and \ref{tab:table4}. In the last column of Table \ref{tab:table4}, the first entry corresponds to $\nu$; the rows below represent orbital parameters: the first row lists the orbital parameters $\left\lbrace \zeta_{1}, \zeta_{2} \right\rbrace$ for $1s\left(\zeta_{1}\right)1s\left(\zeta_{2}\right)$, the second lists $\left\lbrace \zeta_{3}, \zeta_{4} \right\rbrace$ for $2s\left(\zeta_{3}\right)2s\left(\zeta_{4}\right)$, and the third lists $\left\lbrace \zeta_{5}, \zeta_{6} \right\rbrace$ for $2p\left(\zeta_{5}\right)2p\left(\zeta_{6}\right)$.

These tables primarily demonstrate that the chosen basis functions are fewer in number to converge  towards the complete space compared with other functions of sets proposed in the literature \cite{2_Shull_1959, 13_Shull_1955, 35_Slater_1930, 53_Guseinov_2002}.\\
The total energy data reported in Tables \ref{tab:table1} and \ref{tab:table2} correspond to calculations performed using the minimal and double$-$zeta basis set approximations. Specifically, for the minimal basis set approximation Table \ref{tab:table1} presents results for the $\left\lbrace 1+\nu \right\rbrace_{0}$ configuration, denoted as $\left(12\right)$, while Table \ref{tab:table2} shows results for the $\left\lbrace 1+\nu \right\rbrace_{1}$ configuration, denoted $\left(12-2\right)$. Double$-$zeta basis sets are considered, corresponding to the $\left\lbrace \nu \right\rbrace_{0}$, denoted, $\left(1111\right)$ in Table \ref{tab:table1} and $\left\lbrace 1+\nu \right\rbrace_{1}$, denoted, $\left(1122-22\right)$ in Table \ref{tab:table2}, respectively. Results are also provided using exactly the same variational parameters given in Ref. \cite{27_Koga_1995} for the basis set approximation $\left\lbrace 1+1 \right\rbrace_{0}$  corresponding to the orbital configuration $\left(1111222\right)$, 
\begin{align*}
\left\lbrace 1s\left(\zeta_{1}\right), 1s\left(\zeta_{2}\right), 1s\left(\zeta_{3}\right), 1s\left(\zeta_{4}\right), 2s\left(\zeta_{5}\right), 2s\left(\zeta_{6}\right), 2s\left(\zeta_{7}\right)\right\rbrace
\end{align*}
and for $\left\lbrace 2+1 \right\rbrace_{1}$ corresponding to the orbital configuration $\left(1111122-22223\right)$, 
\begin{align*}
\left\lbrace
\begin{array}{l}
1s(\zeta_1), 1s(\zeta_2), 1s(\zeta_3), 1s(\zeta_4), 1s(\zeta_5), 2s(\zeta_6), 2s(\zeta_7), \\
2p(\zeta_8), 2p(\zeta_9), 2p(\zeta_{10}), 2p(\zeta_{11}), 3p(\zeta_{12})
\end{array}
\right\rbrace
\end{align*}
in these tables, respectively. The boldface digits represent values matching those obtained by numerical solution of Hartree$-$Fock equations. The underlined digits denote values that coincide with those obtained through the use of STFs in solution of the Hartree$-$Fock$-$Roothaan equations, employing an identical basis set approximation.

Tables \ref{tab:table1} and \ref{tab:table2} demonstrate the lack of stability associated with treating $\alpha$ as a variational parameter. It is evident that the parameter $\alpha$, employed in the definition of complete orthonormal sets of exponential$-$type orbitals within a weighted Hilbert space, exhibits a systematic dependence on both the nuclear charge and the number of electrons for minimal basis sets. When a double$-$zeta basis set is used, this regularity is not conserved, which may be attributed to the increased complexity of the optimization procedure. Total energy may converge to a local minimum rather than a global minimum, particularly with higher dimensionality introduced by the additional orbital parameters. Consequently, in the minimal basis set, $\alpha$ is treated as a variational parameter. In contrast, for the double$-$zeta basis set, basis sets corresponding to orbital configrations $\left(1111222\right)$ and $\left(1111122-22223\right)$, results are reported for fixed values of $\alpha$, namely $\alpha = 2$ and $\alpha = 2.99999$, to approximate the limit $\alpha \rightarrow 3$. The BH$-$ETOs clearly enhance the quality of the wave function representation, allowing greater accuracy. The results obtained for $\alpha \rightarrow 3$ demonstrate convergence towards those derived from STFs. In the asymptotic regime, $\alpha$ approaches the value $3$. This behavior provides clear evidence that $\alpha$ cannot be regarded as a non-linear variational parameter, contrary to certain assumptions encountered in the literature \cite{54_Guseinov_2012, 55_Guseinov_2014}, where the minimum total energies for certain atoms have been reported precisely at the limiting value $\alpha \rightarrow 3$. An illustrative case is the Li$^-$ atom, as reported in Table \ref{tab:table1}, where the optimization tends to approach the imposed upper limit of $\alpha_{max} = 2.9$.

Whether the HF$-$limit can be achieved or not is tested for Be atom. Five $1s$ and five $2s$ orbitals are used, denoted $\left( 1111122222 \right)$. They define the expansion of the basis. The non$-$integer parameter $\nu$, the weighting parameter $\alpha$ (treated as variational) and orbital parameters obtained through optimization are given, respectively, as follows:
\begin{align*}
\left\lbrace
\begin{array}{l}
1.00001 01316, 0.76334 36854, 1.46634 67859,
\\
3.72163 92141, 14.61000 42129, 7.19503 50726,
\\
1.21100 77606, 2.84430 73746, 5.74513 59725,
\\
9.08802 40131, 3.40775 74368, 0.05350 28025
\end{array}
\right\rbrace
\end{align*}
The computed total energy, $E= -14.57302 31669 99873$ (tabulated value positive -E), agrees with the numerically exact Hartree$–$Fock solution to ten significant digits.
\begin{table*}[h!]
\renewcommand{\arraystretch}{1.3}
\caption{\label{tab:table1}
BH$-$ETO total energies $\left(E\right)$ in atomic units $\left( a.u. \right)$ calculated using two basis set configurations, minimal $(12)$, double$-$zeta $(1111)$ and $\left(1111222\right)$, with variational parameters $\left\lbrace \nu, \zeta_{i}, \alpha \right\rbrace$, for the ground state $\left(^{1}S\right)$ of the $Be$ isoelectronic series and the $Li^{-}$ ion.
}
\begin{threeparttable}
\begin{ruledtabular}
\begin{tabular}{ccccc}
Atom & 
\begin{tabular}[c]{@{}l@{}}
\multicolumn{1}{c}{$E$ for $\left(12\right)$}
\\
\multicolumn{1}{c}{$\left( \nu_{opt}\right.$, $\zeta_{opt}$ and $\left.\alpha_{opt}\right)$}
\end{tabular}
& 
\begin{tabular}[c]{@{}l@{}}
\multicolumn{1}{c}{$E$ for $\left(12\right)$}
\\
\multicolumn{1}{c}{$\left(\zeta_{opt}\right.$, $\left.\alpha \rightarrow 3\Rightarrow n \in \mathbb{N}_{>0}\right)$}
\end{tabular}
&
\begin{tabular}[c]{@{}l@{}}
\multicolumn{1}{c}{$E$ for $\left(1111\right)$}
\\
\multicolumn{1}{c}{$\left(\nu_{opt}\right.$, $\zeta_{opt}$, $\left.\alpha=2\right)$}
\\
\multicolumn{1}{c}{$\left(\zeta_{opt}\right.$, $\left.\alpha \rightarrow 3\Rightarrow n \in \mathbb{N}_{>0}\right)$}
\end{tabular}
&
\begin{tabular}[c]{@{}l@{}}
\multicolumn{1}{c}{$E$ for $\left(1111222\right)$}
\\
\multicolumn{1}{c}{$\left(\zeta\right.$ from \cite{27_Koga_1995}, $\left.\alpha \rightarrow 3\Rightarrow n \in \mathbb{N}_{>0}\right)$}
\end{tabular} 
\\ \cline{2-5}
$Li^{-}$ 
&
\begin{tabular}[c]{@{}l@{}}
\textbf{-7.41}727 35923 83753
\end{tabular}
&
\begin{tabular}[c]{@{}l@{}}
\textbf{-7.4}1080 15873 73588 
\end{tabular}
&
\begin{tabular}[c]{@{}l@{}}
\textbf{-7.427}98 32904 77606 
\\
\textbf{-7.427}97 80212 86749
\end{tabular}
& 
\begin{tabular}[c]{@{}l@{}}
\textbf{-7.42823 2061} \tnote{e}
\end{tabular}
\\ \cline{2-5}
$Be$ 
&
\begin{tabular}[c]{@{}l@{}}
\textbf{-14.5}6492 26469 04510 
\\
\textbf{-14.5}5884 765 \tnote{a}
\end{tabular}
&
\begin{tabular}[c]{@{}l@{}}
\underline{\textbf{-14.5}5673} 72648 99524 
\\
\underline{\textbf{-14.5}5673} 9859 \tnote{b}
\end{tabular}
&
\begin{tabular}[c]{@{}l@{}}
\textbf{-14.5730}1 02840 55339 
\\
\underline{\textbf{-14.5730}0 949}24 44817 
\\
\underline{\textbf{-14.5730}0 95} \tnote{c}
\end{tabular}
& 
\begin{tabular}[c]{@{}l@{}}
\underline{\textbf{-14.57302 31}455} 47331
\\
\underline{\textbf{-14.57302 31}46} \tnote{d}
\\
\textbf{-14.57302 317} \tnote{e}
\end{tabular}
\\ \cline{2-5}
$B^{+}$ 
&
\begin{tabular}[c]{@{}l@{}}
\textbf{-24.23}022 63044 21595 
\end{tabular}
&
\begin{tabular}[c]{@{}l@{}}
\underline{\textbf{-24.2}1357 1}3278 99284 
\\
\underline{\textbf{-24.2}1357 1}151 \tnote{b}
\end{tabular}
&
\begin{tabular}[c]{@{}l@{}}
\textbf{-24.237}08 99808 46127 
\\
\textbf{-24.236}98 32866 28266
\end{tabular}
& 
\begin{tabular}[c]{@{}l@{}}
\underline{\textbf{-24.23757 51755}} 13178
\\
\underline{\textbf{-24.23757 5176}} \tnote{d}
\\
\textbf{-24.23757 518} \tnote{e}
\end{tabular}
\\ \cline{2-5}
$C^{2+}$ 
&
\begin{tabular}[c]{@{}l@{}}
\textbf{-36.40}130 80035 97424 
\end{tabular}
&
\begin{tabular}[c]{@{}l@{}}
\underline{\textbf{-36.3}70}29 85857 81181 
\\
\underline{\textbf{-36.3}70}32 6680 \tnote{b}
\end{tabular}
&
\begin{tabular}[c]{@{}l@{}}
\textbf{-36.408}08 62568 25231 
\\
\textbf{-36.408}00 37297 64007 
\end{tabular}
& 
\begin{tabular}[c]{@{}l@{}}
\underline{\textbf{-36.40849 5318}}3 73559
\\
\underline{\textbf{-36.40849 5318}} \tnote{d}
\\
\textbf{-36.40849 532} \tnote{e}
\end{tabular}
\\ \cline{2-5}
$N^{3+}$ 
&
\begin{tabular}[c]{@{}l@{}}
\textbf{-51.07}558 91384 54587 
\end{tabular}
&
\begin{tabular}[c]{@{}l@{}}
\underline{\textbf{-51.0}239}1 99698 55167 
\\
\underline{\textbf{-51.0}239}4 7966 \tnote{b}
\end{tabular}
&
\begin{tabular}[c]{@{}l@{}}
\textbf{-51.081}96 15692 29371 
\\
\textbf{-51.081}84 73159 76135
\end{tabular}
& 
\begin{tabular}[c]{@{}l@{}}
\underline{\textbf{-51.08231 695}4}1 33534
\\
\underline{\textbf{-51.08231 695}4} \tnote{d}
\\
\textbf{-51.08231 696} \tnote{e}
\end{tabular}
\\ \cline{2-5}
$O^{4+}$ 
&
\begin{tabular}[c]{@{}l@{}}
\textbf{-68.25}111 78805 83673 
\end{tabular}
&
\begin{tabular}[c]{@{}l@{}}
\underline{\textbf{-68.1}73}19 61310 62880 
\\
\underline{\textbf{-68.1}73}21 0015 \tnote{b}
\end{tabular}
&
\begin{tabular}[c]{@{}l@{}}
\textbf{-68.257}43 55265 73289 
\\
\textbf{-68.257}29 33263 95948
\end{tabular}
& 
\begin{tabular}[c]{@{}l@{}}
\underline{\textbf{-68.25771 056}27} 39885 
\\
\underline{\textbf{-68.25771 056}3} \tnote{d}
\\
\textbf{-68.25771 056} \tnote{e} 
\end{tabular}
\\ \cline{2-5}
$F^{5+}$ 
&
\begin{tabular}[c]{@{}l@{}}
\textbf{-87.92}760 49498 84237
\end{tabular}
&
\begin{tabular}[c]{@{}l@{}}
\underline{\textbf{-87}.8175}1 99244 20202 
\\
\underline{\textbf{-87}.8175}2 1525 \tnote{b}
\end{tabular}
&
\begin{tabular}[c]{@{}l@{}}
\textbf{-87.9340}2 14516 39024
\\
\textbf{-87.933}67 36854 22440
\end{tabular}
& 
\begin{tabular}[c]{@{}l@{}}
\underline{\textbf{-87.93405 305}3}1 10154
\\
\underline{\textbf{-87.93405 305}3} \tnote{d}
\\
\textbf{-87.93405 305} \tnote{e}
\end{tabular}

\end{tabular}
\begin{tablenotes}
\item {$\alpha \rightarrow 3$ with $\alpha=2.99999, n^{\ast} \rightarrow \mathbb{N}_{>0}$}
\item[a] Ref. \cite{55_Guseinov_2014} (with $\psi^{\alpha^{\ast}}$ and minimal basis set configuration. $\alpha^{\ast}$ is optimized)
\item[b] Ref. \cite{50_Erturk_2009}  (with STFs and minimal basis set configuration)
\item[c] Ref. \cite{56_Koga_1998} (with STFs and double$-$zeta basis set configuration) 
\item[d] Refs. \cite{27_Koga_1995, 28_Koga_1995} (Hartree$-$Fock-Roothaan total energies with STFs and $1111222$ basis set configuration)
\item[e] Refs. \cite{27_Koga_1995, 28_Koga_1995} (numerical Hartree$-$Fock total energies)
\end{tablenotes}
\end{ruledtabular}
\end{threeparttable}
\end{table*}
\begin{table*}[h!]
\renewcommand{\arraystretch}{1.3}
\caption{\label{tab:table2}
BH$-$ETO total energies $\left(E\right)$ in atomic units $\left( a.u. \right)$ calculated using two basis set configurations, minimal $(12-2)$, double$-$zeta $(1122-22)$ and $\left(1111122-22223\right)$ with variational parameters $\left\lbrace \nu, \zeta_{i}, \alpha \right\rbrace$, for the ground state $\left(^{1}{S}\right)$ of the $Ne$ isoelectronic series and the $F^{-}$ anion.
}
\begin{threeparttable}
\begin{ruledtabular}
\begin{tabular}{ccccc}
Atom & 
\begin{tabular}[c]{@{}l@{}}
\multicolumn{1}{c}{$E$ for $\left(12-2\right)$}
\\
\multicolumn{1}{c}{$\left(\nu_{opt}\right.$, $\zeta_{opt}$ and $\left.\alpha_{opt}\right)$}
\end{tabular}
& 
\begin{tabular}[c]{@{}l@{}}
\multicolumn{1}{c}{$E$ for $\left(12-2\right)$}
\\
\multicolumn{1}{c}{$\left(\zeta_{opt}\right.$, $\left.\alpha \rightarrow 3\Rightarrow n \in \mathbb{N}_{>0}\right)$}
\end{tabular}
&
\begin{tabular}[c]{@{}l@{}}
\multicolumn{1}{c}{$E$ for $\left(1122-22\right)$}
\\
\multicolumn{1}{c}{$\left(\nu_{opt}\right.$, $\zeta_{opt}$ and $\left.\alpha=2\right)$}
\end{tabular}
&
\begin{tabular}[c]{@{}l@{}}
\multicolumn{1}{c}{$E$ for $\left(1111122-22223\right)$}
\\
\multicolumn{1}{c}{$\left(\zeta\right.$ from \cite{27_Koga_1995}, $\left. \alpha \rightarrow 3\Rightarrow n \in \mathbb{N}_{>0}\right)$}
\end{tabular}
\\ \hline
$F^{-}$ 
& 
\begin{tabular}[c]{@{}l@{}}
\textbf{-98}.72427 26521 63216 
\end{tabular}
&
\begin{tabular}[c]{@{}l@{}}
\textbf{-98}.69650 45053 01074 
\end{tabular}
& 
\begin{tabular}[c]{@{}l@{}}
\textbf{-99.4}3807 79543 53993 
\end{tabular}
& 
\begin{tabular}[c]{@{}l@{}}
\textbf{-99.45945 391} \tnote{e}
\end{tabular}
\\ \cline{2-5}
$Ne$ 
& 
\begin{tabular}[c]{@{}l@{}}
\textbf{-127}.84859 73781 15539 
\\
\textbf{-127}.81732 229 \tnote{a}
\end{tabular}
&
\begin{tabular}[c]{@{}l@{}}
\underline{\textbf{-127}.81218} 11153 75182 
\\
\underline{\textbf{-127}.81218} 0947 \tnote{b}
\end{tabular}
&
\begin{tabular}[c]{@{}l@{}}
\textbf{-128.53}598 17128 87771
\\
\textbf{-128.53}586 6 \tnote{c}
\end{tabular}
&
\begin{tabular}[c]{@{}l@{}}
\textbf{\underline{-128.5470}}2 85842 61361
\\
\textbf{\underline{-128.5470}9 7}973 \tnote{d}
\\
\textbf{-128.54709 81} \tnote{e}
\end{tabular}
\\ \cline{2-5}
$Na^{+}$ 
& 
\begin{tabular}[c]{@{}l@{}}
\textbf{-160}.99799 77481 33933 
\end{tabular}
&
\begin{tabular}[c]{@{}l@{}}
\underline{\textbf{-160}.94802} 48963 61785 
\\
\underline{\textbf{-160}.94802} 5909 \tnote{b}
\end{tabular}
&
\begin{tabular}[c]{@{}l@{}}
\textbf{-161.6}4789 19593 63061 
\end{tabular}
&
\begin{tabular}[c]{@{}l@{}}
\textbf{\underline{-161.6769}}2 73009 37601
\\
\textbf{\underline{-161.6769}6 2}467 \tnote{d}
\\
\textbf{-161.67696 26} \tnote{e}
\end{tabular}
\\ \cline{2-5}
$Mg^{2+}$ 
& 
\begin{tabular}[c]{@{}l@{}}
\textbf{-198}.16412 25110 09889 
\end{tabular}
&
\begin{tabular}[c]{@{}l@{}}
\underline{\textbf{-198}.09525} 31067 36642 
\\
\underline{\textbf{-198}.09525} 4997 \tnote{b}
\end{tabular}
&
\begin{tabular}[c]{@{}l@{}}
\textbf{-198.8}2433 56958 33906 
\end{tabular}
&
\begin{tabular}[c]{@{}l@{}}
\textbf{\underline{-198.830}7}9 74706 55784
\\
\textbf{\underline{-198.830}81 0}154 \tnote{d}
\\
\textbf{-198.83081 04} \tnote{e}
\end{tabular}
\\ \cline{2-5}
$Al^{3+}$ 
& 
\begin{tabular}[c]{@{}l@{}}
\textbf{-239}.34210 99575 22552 
\end{tabular}
&
\begin{tabular}[c]{@{}l@{}}
\underline{\textbf{-23}9.248}44 34659 04755 
\\
\underline{\textbf{-23}9.248}87 2831 \tnote{b}
\end{tabular}
&
\begin{tabular}[c]{@{}l@{}}
\textbf{-239.9}4256 24334 18995 
\end{tabular}
&
\begin{tabular}[c]{@{}l@{}}
\textbf{\underline{-240.00034}} 54704 72285
\\
\textbf{\underline{-240.00034} 8}279
\\
\textbf{-240.00034 86} \tnote{e}
\end{tabular}
\\ \cline{2-5}
$Si^{4+}$ 
& 
\begin{tabular}[c]{@{}l@{}}
\textbf{-284}.52909 68040 18849 
\end{tabular}
&
\begin{tabular}[c]{@{}l@{}}
\underline{\textbf{-28}4.405}52 16370 01334 
\\
\underline{\textbf{-28}4.405}71 4363 \tnote{b}
\end{tabular}
&
\begin{tabular}[c]{@{}l@{}}
\textbf{-285.1}6312 91180 75236 
\end{tabular}
&
\begin{tabular}[c]{@{}l@{}}
\textbf{\underline{-285.18093} 0}8178 81852
\\
\textbf{\underline{-285.18093} 1}065 \tnote{d}
\\
\textbf{-285.18093 14} \tnote{e}
\end{tabular}
\\ \cline{2-5}
$P^{5+}$ 
& 
\begin{tabular}[c]{@{}l@{}}
\textbf{-333}.72265 76739 72560 
\end{tabular}
&
\begin{tabular}[c]{@{}l@{}}
\underline{\textbf{-333}.56}200 60263 25598 
\\
\underline{\textbf{-333}.56}364 1572 \tnote{b}
\end{tabular}
&
\begin{tabular}[c]{@{}l@{}}
\textbf{-334.2}5557 20683 75186 
\end{tabular}
&
\begin{tabular}[c]{@{}l@{}}
\textbf{\underline{-334.36966 0}}1922 35378
\\
\textbf{\underline{-334.36966 0}}364 \tnote{d}
\\
\textbf{-334.36966 07} \tnote{e}
\end{tabular}

\end{tabular}
\begin{tablenotes}
\item {$\alpha \rightarrow 3$ with $\alpha=2.99999$}
\item[a] Ref. \cite{55_Guseinov_2014} (minimal basis set with $\psi^{\alpha^{\ast}}$, $\alpha^{\ast}$ optimized)
\item[b] Ref. \cite{50_Erturk_2009} (minimal basis set with STFs)
\item[c] Ref. \cite{56_Koga_1998} (double$-$zeta basis set with STFs) 
\item[d] Refs. \cite{27_Koga_1995, 28_Koga_1995} (Hartree$-$Fock$-$Roothaan total energies with STFs and $\left(1111122-22223\right)$ basis set configuration)
\item[d] Refs. \cite{27_Koga_1995, 28_Koga_1995} (numerical Hartree$-$Fock total energies)
\end{tablenotes}
\end{ruledtabular}
\end{threeparttable}
\end{table*}
\begin{table*}[h!]
\renewcommand{\arraystretch}{1.15}
\caption{\label{tab:table3}
Optimized variational parameters for results given in Table \ref{tab:table1} with minimal $\left(12\right)$ and double$-$zeta $\left(1111\right)$ basis set approximations.
}
\begin{threeparttable}
\begin{ruledtabular}
\begin{tabular}{ccccc}
Atom &  
$\left( \nu_{opt}, \zeta_{opt},\alpha_{opt} \right)$ for $\left(12\right)$ & 
$\left( \zeta_{opt} \right)$ for $\left(12\right)$ and $\alpha \rightarrow 3$&
$\left( \nu_{opt}, \zeta_{opt} \right)$ for $\left(1111\right)$&
$\left( \zeta_{opt} \right)$ for $\left(1111\right)$ 
\\ \cline{2-5}
$Li^{-}$ 
&
\begin{tabular}[c]{@{}l@{}}
0.97011 63825
\\
2.60496 97070
\\ 
0.47929 72405
\\ 
2.89990 41450 
\end{tabular}
&
\begin{tabular}[c]{@{}l@{}}
2.68881 75678 
\\
0.48528 13454 
\end{tabular}
&
\begin{tabular}[c]{@{}l@{}}
0.99979 84887 
\\
4.55769 70490 
\\
2.45693 35946
\\
1.72646 26387
\\
0.28865 29270
\end{tabular}
& 
\begin{tabular}[c]{@{}l@{}}
4.44580 60567 
\\
2.42637 93834 
\\ 
1.78469 83545
\\ 
0.28751 40144
\end{tabular}
\\ \cline{2-5}
$Be$ 
&
\begin{tabular}[c]{@{}l@{}}
0.97956 747171
\\
3.61314 46496
\\ 
1.00720 23726 
\\ 
2.49046 02359
\end{tabular}
&
\begin{tabular}[c]{@{}l@{}}
3.68311 95179
\\
0.95612 16976
\end{tabular}
&
\begin{tabular}[c]{@{}l@{}}
1.00041 74856
\\
6.44517 73995 
\\
3.47272 35581
\\
1.77903 99021
\\
0.72614 27913
\end{tabular}
& 
\begin{tabular}[c]{@{}l@{}}
6.37553 30417
\\
3.46642 42495 
\\ 
1.77786 10618 
\\ 
0.72615 48576 
\end{tabular}
\\ \cline{2-5}
$B^{+}$ 
&
\begin{tabular}[c]{@{}l@{}}
0.98485 83007 
\\
4.62254 94216 
\\ 
1.52396 24414 
\\
2.18571 07121 
\end{tabular}
&
\begin{tabular}[c]{@{}l@{}}
4.67601 97952 
\\
1.39705 33051 
\end{tabular}
&
\begin{tabular}[c]{@{}l@{}}
0.99725 10695 
\\
6.26337 82967 
\\
1.32875 33885 
\\
4.10808 45371 
\\
1.72542 83209 
\end{tabular}
& 
\begin{tabular}[c]{@{}l@{}}
6.86830 53745 
\\
1.44513 03795 
\\ 
4.28019 40690 
\\
1.54735 24255 
\end{tabular}
\\ \cline{2-5}
$C^{2+}$ 
&
\begin{tabular}[c]{@{}l@{}}
0.98899 84844 
\\
5.63791 22542 
\\ 
2.02298 93324 
\\ 
2.03821 51003 
\end{tabular}
&
\begin{tabular}[c]{@{}l@{}}
5.66458 43145 
\\
1.83433 64075 
\end{tabular}
&
\begin{tabular}[c]{@{}l@{}}
0.99666 90825 
\\
6.64382 70104 
\\
4.58903 43866 
\\
2.79319 88859 
\\
1.62489 73212 
\end{tabular}
& 
\begin{tabular}[c]{@{}l@{}}
7.03694 79464 
\\
4.80639 59667 
\\ 
2.62990 38347 
\\ 
1.64405 33298 
\end{tabular}
\\ \cline{2-5}
$N^{3+}$ 
&
\begin{tabular}[c]{@{}l@{}}
0.98913 69406 
\\
6.62038 02997 
\\ 
2.53711 97768 
\\ 
1.85135 53935 
\end{tabular}
&
\begin{tabular}[c]{@{}l@{}}
6.65221 71589 
\\
2.25472 70750 
\end{tabular}
&
\begin{tabular}[c]{@{}l@{}}
0.99805 30736 
\\
7.92609 68182 
\\
5.65724 47500 
\\
3.20697 05224 
\\
2.10917 16741 
\end{tabular}
& 
\begin{tabular}[c]{@{}l@{}}
8.35995 94588 
\\
5.75253 83192 
\\ 
3.33524 87831 
\\ 
2.11795 75655 
\end{tabular}
\\ \cline{2-5}
$O^{4+}$ 
&
\begin{tabular}[c]{@{}l@{}}
0.99060 55392 
\\
7.62068 33268 
\\ 
3.04047 28509 
\\ 
1.74679 09290 
\end{tabular}
&
\begin{tabular}[c]{@{}l@{}}
7.63976 16246 
\\
2.68371 61249 
\end{tabular}
&
\begin{tabular}[c]{@{}l@{}}
0.99899 34502 
\\
9.33699 76476 
\\
6.70141 57153 
\\
3.85124 90279 
\\
2.55008 45672 
\end{tabular}
& 
\begin{tabular}[c]{@{}l@{}}
9.25392 43380 
\\
6.62749 57584 
\\ 
3.74064 27966 
\\ 
2.57695 26322 
\end{tabular}
\\ \cline{2-5}
$F^{5+}$ 
&
\begin{tabular}[c]{@{}l@{}}
0.99173 83317 
\\
8.62158 56677 
\\
3.53694 60431 
\\
1.68297 48950 
\end{tabular}
&
\begin{tabular}[c]{@{}l@{}}
8.62707 47166 
\\
3.11218 42717 
\end{tabular}
&
\begin{tabular}[c]{@{}l@{}}
1.00005 57108 
\\
13.99983 47326
\\
8.39501 74457 
\\
4.30049 18266 
\\
3.03589 72070 
\end{tabular}
& 
\begin{tabular}[c]{@{}l@{}}
10.66642 35397 
\\
7.49985 12593 
\\ 
4.21544 87717 
\\ 
3.07275 28559 
\end{tabular}

\end{tabular}
\begin{tablenotes}
\item {$\alpha \rightarrow 3$ with $\alpha=2.99999, n^{\ast} \rightarrow \mathbb{N}_{>0}$}
\end{tablenotes}
\end{ruledtabular}
\end{threeparttable}
\end{table*}
\begin{table*}[h!]
\renewcommand{\arraystretch}{1.15}
\caption{\label{tab:table4}
Optimized variational parameters for results given in Table \ref{tab:table2} with minimal $\left(12-2\right)$ and double$-$zeta $\left(1122-22\right)$ basis set approximations.
}
\begin{threeparttable}
\begin{ruledtabular}
\begin{tabular}{cccc}
Atom & 
$\left( \nu_{opt}, \zeta_{opt},\alpha_{opt} \right)$  for $\left(12-2\right)$ & 
$\left( \zeta_{opt} \right)$ for $\left(12-2\right)$ and $\alpha \rightarrow 3$ &
$\left( \nu_{opt}, \zeta_{opt} \right)$ for $\left(1122-22\right)$ and $\alpha=2$
\\ \hline
$F^{-}$ 
& 
\begin{tabular}[c]{@{}l@{}}
0.98226 06280
\\
8.49952 61354
\\
2.52974 19409
\\
2.32501 02604
\\
2.77178 48831
\end{tabular}
&
\begin{tabular}[c]{@{}l@{}}
8.65706 32990 
\\
2.49313 81677
\\
2.34423 54210
\end{tabular}
& 
\begin{tabular}[c]{ccc}
\multicolumn{3}{c}{0.99687 47986}
\\
\begin{tabular}[c]{@{}l@{}}
1.75468 49714
\\
5.88135 41022
\end{tabular}
& 
\begin{tabular}[c]{@{}l@{}}
6.75024 99772  
\\
9.50855 45706
\end{tabular}
&
\begin{tabular}[c]{@{}l@{}}
3.83747 76750 
\\
1.51730 18282
\end{tabular}
\end{tabular}
\\ \cline{2-4}
$Ne$ 
& 
\begin{tabular}[c]{@{}l@{}}
0.98253 39174 
\\
9.47334 28896 
\\
2.97948 86577 
\\
2.85474 15082 
\end{tabular}
&
\begin{tabular}[c]{@{}l@{}}
9.64226 56782 
\\
2.87934 95290 
\\
2.87902 80249 
\end{tabular}
&
\begin{tabular}[c]{ccc}
\multicolumn{3}{c}{0.99825 93378}
\\
\begin{tabular}[c]{@{}l@{}}
9.72712 42395
\\
2.06189 04534
\end{tabular}
& 
\begin{tabular}[c]{@{}l@{}}
7.94344 90396 
\\
13.16481 86693
\end{tabular}
&
\begin{tabular}[c]{@{}l@{}}
4.67024 88162
\\
2.05089 38267
\end{tabular}
\end{tabular}
\\ \cline{2-4}
$Na^{+}$ 
& 
\begin{tabular}[c]{@{}l@{}}
0.98330 15839 
\\
10.45735 83937 
\\
3.45449 83769 
\\
3.37451 98391 
\\
2.45872 60866 
\end{tabular}
&
\begin{tabular}[c]{@{}l@{}}
10.62575 77291 
\\
3.28053 07120 
\\
3.40250 74345 
\end{tabular}
&
\begin{tabular}[c]{ccc}
\multicolumn{3}{c}{0.99501 46384}
\\
\begin{tabular}[c]{@{}l@{}}
9.30567 11417 
\\
2.45056 57837
\end{tabular}
& 
\begin{tabular}[c]{@{}l@{}}
10.66468 23365 
\\
7.72449 54605
\end{tabular}
&
\begin{tabular}[c]{@{}l@{}}
2.25323 93242 
\\
4.83974 42843 
\end{tabular}
\end{tabular}
\\ \cline{2-4}
$Mg^{2+}$ 
& 
\begin{tabular}[c]{@{}l@{}}
0.98375 82684 
\\
11.43525 99053 
\\
3.94050 95523 
\\
3.88797 11275 
\\
2.32976 08824 
\end{tabular}
&
\begin{tabular}[c]{@{}l@{}}
11.61028 25640 
\\
3.68920 18603 
\\
3.92010 44426 
\end{tabular}
&
\begin{tabular}[c]{ccc}
\multicolumn{3}{c}{0.99812 79176}
\\
\begin{tabular}[c]{@{}l@{}}
8.63134 56480
\\
2.83332 13164
\end{tabular}
& 
\begin{tabular}[c]{@{}l@{}}
8.45017 58064
\\
11.14474 66879
\end{tabular}
&
\begin{tabular}[c]{@{}l@{}}
3.04185 10394
\\
6.14112 23457
\end{tabular}
\end{tabular}
\\ \cline{2-4}
$Al^{3+}$ 
& 
\begin{tabular}[c]{@{}l@{}}
0.98450 51322
\\
12.42902 84003
\\
4.44108 54296
\\
4.39813 29108
\\
2.21012 55711
\end{tabular}
&
\begin{tabular}[c]{@{}l@{}}
12.61591 02418
\\
4.10101 09078
\\
4.43458 55008
\end{tabular}
&
\begin{tabular}[c]{ccc}
\multicolumn{3}{c}{0.99230 97835}
\\
\begin{tabular}[c]{@{}l@{}}
15.67032 65866
\\
3.24255 11727
\end{tabular}
& 
\begin{tabular}[c]{@{}l@{}}
9.50374 81035
\\
14.34860 34355
\end{tabular}
&
\begin{tabular}[c]{@{}l@{}}
5.79149 15716
\\
3.27206 08325
\end{tabular}
\end{tabular}
\\ \cline{2-4}
$Si^{4+}$ 
& 
\begin{tabular}[c]{@{}l@{}}
0.98559 92166
\\
13.42153 77457
\\
4.91570 74391
\\
4.90818 58887
\\
2.14447 38860
\end{tabular}
&
\begin{tabular}[c]{@{}l@{}}
13.57402 14130
\\
4.51990 81014
\\
4.93762 72806
\end{tabular}
&
\begin{tabular}[c]{ccc}
\multicolumn{3}{c}{0.99640 91871}
\\
\begin{tabular}[c]{@{}l@{}}
7.37486 47781
\\
3.71190 29269
\end{tabular}
& 
\begin{tabular}[c]{@{}l@{}}
8.94654 40937
\\
13.11789 09915
\end{tabular}
&
\begin{tabular}[c]{@{}l@{}}
6.94206 10108
\\
3.82752 44614
\end{tabular}
\end{tabular}
\\ \cline{2-4}
$P^{5+}$ 
& 
\begin{tabular}[c]{@{}l@{}}
0.98625 57929
\\
14.41276 28023
\\
5.41432 61889
\\
5.41550 68635
\\
2.06190 65963
\end{tabular}
&
\begin{tabular}[c]{@{}l@{}}
14.54202 31886
\\
4.95258 34019
\\
5.43416 76062
\end{tabular}
&
\begin{tabular}[c]{ccc}
\multicolumn{3}{c}{0.99803 22501}
\\
\begin{tabular}[c]{@{}l@{}}
14.95028 45306
\\
3.83622 27840 
\end{tabular}
& 
\begin{tabular}[c]{@{}l@{}}
14.86465 86715
\\
11.87948 71225
\end{tabular}
&
\begin{tabular}[c]{@{}l@{}}
11.87243 13940
\\
5.16841 66250 
\end{tabular}
\end{tabular}

\end{tabular}
\begin{tablenotes}
\item {$\alpha \rightarrow 3$ with $\alpha=2.99999, n^{\ast} \rightarrow \mathbb{N}_{>0}$}
\end{tablenotes}
\end{ruledtabular}
\end{threeparttable}
\end{table*}
\twocolumngrid
\section*{Conclusion}
The basis used in this work is orthonormal and complete (from Sturm$-$Liouville theory). It is in fact obtained in by taking solutions of the Dirac equation (a complete set of ket eigenfunctions spanning a square integrable Hilbert space), to the non-relativistic limit.  

It is based on solving the Kepler problem defined by Infeld and Hull \cite{10_Infeld_1951}, it is beyond standard Hilbert space which is a subspace of the present basis space. The eigenfunctions obtained can be thought of as spanning the space of Coulomb$-$Sturmians  or Lambda or Guseinov functions \cite{53_Guseinov_2002} with general weight.

It turns out that the previously published studies (notably on Slater Type functions), which were thought to be on firm ground have been completely superseded by the present work. Furthermore, in the case of Slater$-$type functions it has proven difficult to improve Clementi, Roetti \cite{29_Clementi_1974} and Bunge et al. \cite{30_Bunge_1993} screening constants. Some attempts to do so fail to take into account liner dependence they introduce into a Slater basis during variational optimization carried out globally.

\section*{Acknowledgement}
The authors declares no conflict of interest.
\pagebreak

\end{document}